\documentclass[conference,a4paper]{IEEEtran}
\usepackage{cite}
\usepackage{amsmath,amssymb,amsfonts}
\usepackage{algorithmic}
\usepackage{graphicx}
\usepackage{textcomp}
\usepackage[caption=false]{subfig}
\usepackage{booktabs}
\usepackage{multirow}
\usepackage{xcolor}

\usepackage{pgfplots}
\usepgfplotslibrary{groupplots,dateplot}
\usepackage{tikz}
\usetikzlibrary{shapes.geometric}
\usetikzlibrary{arrows.meta,arrows}
\usetikzlibrary{patterns,shapes.arrows}
\pgfplotsset{compat=newest}
\usepackage{environ}
\usepackage{multirow}
\usepackage{booktabs}
\usepackage{colortbl}
\usepackage{url}

\usepackage{hyperref}
\usepackage{xcolor}
\definecolor{dark-red}{rgb}{0.4,0.15,0.15}
\definecolor{dark-blue}{rgb}{0.15,0.15,0.8}
\definecolor{medium-blue}{rgb}{0,0,0.5}
\hypersetup{
    colorlinks, linkcolor={dark-red},
    citecolor={dark-blue}, urlcolor={medium-blue}
}

\definecolor{correction}{RGB}{255, 0, 0}

\definecolor{black}{RGB}{0, 0, 0}
\definecolor{orange}{RGB}{230, 159, 0} 
\definecolor{skyblue}{RGB}{86, 180, 233}
\definecolor{green}{RGB}{0, 158, 115} 
\definecolor{yellow}{RGB}{240, 228, 66}
\definecolor{blue}{RGB}{0, 114, 178} 
\definecolor{red}{RGB}{213, 94, 0}
\definecolor{purple}{RGB}{204, 121, 167}
\definecolor{lightgray}{RGB}{204,204,204} 

\def\BibTeX{{\rm B\kern-.05em{\sc i\kern-.025em b}\kern-.08em
    T\kern-.1667em\lower.7ex\hbox{E}\kern-.125emX}}
\begin{document}

\title{Embedded Deep Learning for Bio-hybrid Plant Sensors to Detect Increased Heat and Ozone Levels}

\author{
    Till Aust\IEEEauthorrefmark{1}, 
    Christoph Karl Heck\IEEEauthorrefmark{1}, 
    Eduard Buss, and 
    Heiko Hamann\\
    Department of Computer and Information Science, University of Konstanz, Konstanz, Germany \\
    till.aust@uni-konstanz.de\\
\IEEEauthorblockA{\IEEEauthorrefmark{1} These authors contributed equally}
} 

\maketitle

\begin{abstract}
We present a bio-hybrid environmental sensor system that integrates natural plants and embedded deep learning for real-time, on-device detection of temperature and ozone level changes.
Our system, based on the low-power \textit{PhytoNode} platform, records electric differential potential signals from \textit{Hedera helix} and processes them onboard using an embedded deep learning model. 
We demonstrate that our sensing device detects changes in temperature and ozone with good sensitivity of up to 0.98. 
Daily and inter-plant variability, as well as limited precision, could be mitigated by incorporating additional training data, which is readily integrable in our data-driven framework. 
Our approach also has potential to scale to new environmental factors and plant species.
By integrating embedded deep learning onboard our biological sensing device, we offer a new, low-power solution for continuous environmental monitoring and potentially other fields of application.
\end{abstract}

\begin{IEEEkeywords}
Phytosensing, Bio-hybrid system, Embedded deep learning, Natural plants
\end{IEEEkeywords}

\section{Introduction}
\noindent In recent years phytosensing emerged as a new paradigm on indirectly sensing the environment through natural plants, 
by measuring and analyzing the natural plant's physiological response to their surrounding environment. 
Plant-wearables (i.e., plant sensors monitoring physiological responses) can enable natural plants to act as bio-hybrid sensors. 
Due to their environmental integration, a single plant bio-hybrid sensor can simultaneously detect a wide range of hazards~\cite{Pfotenhauer2024}. 
Although phytosensing is still emerging and less accurate than specialized sensors, it holds significant potential for future applications where sensor density, coverage, and cost-efficiency outweigh the need for high precision, for example, early warning systems in precision agriculture~\cite{Tran2024,Afif2025}, environmental monitoring~\cite{Buss2025}, or air pollution monitoring~\cite{Garcia-Carmona2021}. 

In our project \textit{WatchPlant}~\cite{Garcia-Carmona2021}, we use phytosensing bio-hybrid sensors to monitor urban air pollution, which poses serious health threats~\cite{Drew2025}.
Due to spatial sparsity of existing monitoring solutions, new monitoring systems are developed~\cite{Kumar2015}. 
Our phytosensing-based approach~\cite{Buss2022,Buss2025} offers a cost-effective, eco-friendly alternative by leveraging urban vegetation 
potentially enhancing public acceptance over conventional monitoring stations.

Our bio-hybrid sensor combines a plant (here \textit{Hedera helix} known as ivy) with the \textit{PhytoNode}~\cite{Buss2022,Buss2025}, which records macroscopic plant electrophysiology. 
Other plant-wearables use piezoresistive, chemical, photodetector, or capcaitive sensors~\cite{Park2025}, but can only monitor few environmental changes. 
%
In contrast, measuring plant electrophysiology (electric differential potential~(EDP) in plant tissue, in millivolts) allows for detecting diverse environmental parameters, such as light~\cite{Gurovich2009}, heat\cite{Buss2023}, humidity~\cite{Sun2023}, CO$_2$-levels~\cite{Volkov2012}, O$_3$-levels~\cite{Dolfi2015,Aust2025}, herbivory~\cite{Pachu2023}, salt stress~\cite{Choi2014}, or nutrition deficits~\cite{Tran2024}. 
Existing plant electrophysiology sensors, such as those by \emph{Vivent}~\cite{Vivent} and \emph{CYBRES}~\cite{Cybres}, are closed-source, non-standalone, and not suited for (energy-)autonomous outdoor deployment (e.g., $5$~V, $0.3$~A for CYBRES), unlike the \textit{PhytoNode}~\cite{Buss2025} (e.g., $3.3$~V, $4$~mA).

The \textit{PhytoNode}, exhibits an adaptive, nonlinear response that deviates from classical linear transfer functions. 
Instead, we model its behavior as a classification problem: inferring environmental conditions from patterns in the macroscopic plant EDP; a challenging approach given sophisticated macroscopic plant models that could serve as white-box models are not available in plant biology~\cite{Sukhova2017}.
In absence of white-box models, gray-box models~\cite{Dolfi2015} offer limited adaptability due to their rigid structure. 
Consequently, many studies adopt black-box or deep learning (DL) approaches, analyzing plant EDPs offline.
Onboard classification of plant EDPs remains underexplored due to onboard processing challenges. 
Our study addresses this gap by developing a phytosensing device that facilitates in situ detection of the changing environmental factors temperature or ozone concentration.
We propose a DL~approach for its flexibility embedded on the \textit{PhytoNode}, which poses challenges on memory, processing power, and energy constraints.
Fully convolutional networks (FCNs)~\cite{Wang2017} are parameter-efficient through weight sharing and can effectively classify EDPs~\cite{Buss2023}, whereas alternative architectures or larger models would require more resources for additional parameters and operations.
We previously demonstrated DL feasibility on the \textit{PhytoNode}~\cite{Buss2022} and developed Mbed~Torch~Fusion~OS~\cite{Heck2025b} to flexibly integrate PyTorch models via ExecuTorch~\cite{Executorch}. 
We extend the framework by convolutional layers to deploy embedded FCNs on the \textit{PhytoNode}, enabling onboard classification of temperature and ozone variations, advancing the sensor system toward autonomous environmental monitoring.

\section{Experiments and Methodology}

\noindent To train and test our bio-hybrid sensor, we built two indoor experimental setups, presented in Sec.~\ref{sec:experimental_setup}. 
Our bio-hybrid sensor does not directly measure an interpretable physical quantity.
Instead, it relies on the plant's complex internal processing to recognize environmental conditions. 
The measurement chain starts with collecting raw plant EDPs, which are online preprocessed (Sec.~\ref{sec:preprocessing}), further processed using embedded DL, and finally interpreted (Sec.~\ref{sec:deep_learning}).

\subsection{Experimental setups and protocol}
\label{sec:experimental_setup}
\begin{figure}[t]
    \centering
    \includegraphics[]{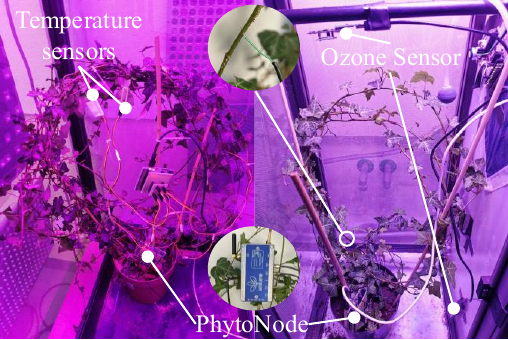}
    \vspace{-10pt} 
    \caption{(Left)~heat stimulus setup, (right)~ozone stimulus setup (compare~\cite{Buss2025}). The upper circle shows how the needle is attached to the stem, while the lower circle depicts the weather-proof \textit{PhytoNode}.}
    \vspace{-15pt} 
    \label{fig:experimental_setup}
\end{figure}
\noindent Our two experimental setups are summarized in Fig.~\ref{fig:experimental_setup}. 
Each consists of an ivy and one \textit{PhytoNode} measuring its EDPs.
The \textit{PhytoNode} has four electrodes measuring the  plant's EDP in two channels on different branches, spanning $\sim$150~cm from near a leaf (at the petiole or stem) down to the stem just above the soil (sampling frequency 100~Hz). 
The whole setup is shielded by a Faraday cage to suppress external electromagnetic interference. 
A~Raspberry~Pi controls the stimuli (e.g., heater or ozone generator) outside the Faraday cage following a fixed protocol and serves as data sink. 
The heat stimulus ($\sim$6~$^{\circ}$C increase, average measure of two DS18B20 temperature sensors in the middle of the setup) is applied for 30~min, five times daily (8:00–20:30) with 2-hour recovery intervals; the ozone stimulus ($\sim$1400~parts per billion (ppb), average of two DGS-O3 968-042 ozone sensors; placed bottom and top of the setup) follows the same protocol. 
To isolate the plant's response, we vary only the target stimulus, while keeping all other conditions constant. 
For training, we did 151~heat stimulus experiments (simultaneously measuring two ivy plants in parallel, total of~8 different plants), resulting in 604~response recordings. 
For training the ozone model we use the data of~\cite{Aust2025} and noise-based data augmentation~\cite{tsaug} to increase data variability. 
For testing, we use 2 new ivies over 5~days for heat and 2 plants for 2~days for the ozone stimulus.

\subsection{Online preprocessing for plant electrophysiology}
\label{sec:preprocessing}
\noindent To enable online classification, preprocessing and inference must run efficiently on-device, meeting microcontroller constrains. 
For both \textit{PhytoNode} channels, we do the same processing.
Initially, we downsample the signal by averaging all EDP readings over six-second intervals. 
This smooths the signal and reduces later processing overhead, while keeping all relevant information, as empirical observations show that meaningful variations occur at timescales of seconds. 
Next, the smoothed signal is converted to millivolts using a constant scaling factor and offset of the ADC.
Every 100~values (10~min slice), we apply one of the following scaling approaches to compensate for inter-plant variability and long-term drift in the EDP:
\begin{figure*}[t]
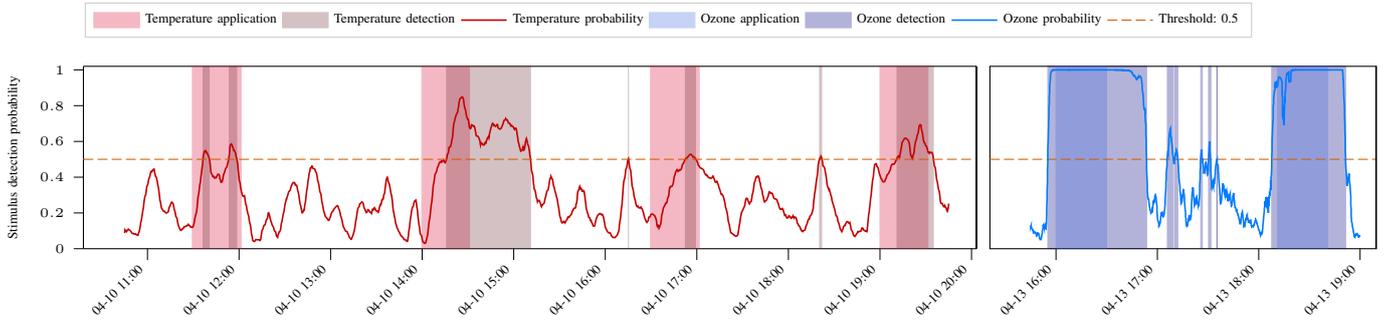

\centering
  \include{figures/live_plot_temperature}
  \vspace{-35pt} 
  \caption{We show the real-time online detection probability for a given stimulus (left, red heat stimulus; right, blue ozone stimulus) taking the minimum detection probability of the two EDP channels connected to one ivy plant.}
  \vspace{-15pt} 
  \label{fig:online_results}
\end{figure*}
\subsubsection*{Adjusted min-max (AMM)}
The EDP is not exceeding $\pm$~200~mV (empirical findings from previous experiments of continuous measurements over multiple weeks). 
We use $x_{\text{AMM}} = (10^3x - g_{\text{min}})/(g_{\text{max}} - g_{\text{min}})$, with sample~$x$, $g_{\text{min}}=-200$~mV, and $g_{\text{max}}=200$~mV. 
Based on empirical observations, we scale the signal by $10^3$ to amplify subtle variations, which have proven beneficial for model training.

\subsubsection*{Scaled min-max (SMM)}
We scale $x$ as $x_{\text{SMM}} = 10^3(x-x_{\text{min}})/(x_{\text{max}}-x_{\text{min}})$ using the current \textit{min} and \textit{max} value of~$x$.

\subsubsection*{Raw data scaled (RDS)}
We scale by~$10^3$: $x_{\text{RDS}}=10^3x$.

\noindent The signal normalized by either method is used for training and classification of the external stimuli (e.g., increased temperature or high levels of ozone). 
From the data we derived balanced training datasets of $5\times 10^4$~samples for heat and about $5\times 10^4$ samples for ozone using window slicing with a stride of 10~seconds (data are available at~\cite{Heck2025c}).

\subsection{Embedded deep learning for plant electrophysiology}
\label{sec:deep_learning}

\noindent For on-device deployment on the \textit{PhytoNode}, we employ an FCN with 3~ReLU-activated convolutional blocks on the STM32WB55RG microcontroller, chosen for its energy efficiency and wireless communication capabilities (c.f.~\cite{Buss2025}).
The FCN is implemented and trained with PyTorch Lightning~\cite{Falcon2019}, and its hyperparameters are optimized using Optuna~\cite{Akiba2019}. 
For each block, we vary channels and kernel size, as well as the learning rate, batch size, and weight decay.
We validate model compatibility with device limits before each optimization round; if memory usage exceeds 130~kB, an alternative configuration is selected and re-validated. 
Models are trained with Adam, exported to ExecuTorch, and deployed on the \textit{PhytoNode} using Mbed~Torch~Fusion~OS.

We train a binary classification model separately for each stimulus (increased temperature, high levels of ozone), and select the appropriate model for each experiment. 
For details of the training process, see our code repository~\cite{Heck2025}.

\section{Results and Discussion}
\label{sec:results}

\noindent We trained and selected models on collected data (Sec.~\ref{sec:model_selection}) and deployed the best for onboard, online classification for increased temperature and ozone levels (Sec.~\ref{sec:online_classification}).
We balanced the training data by sampling to simplify training and allowing for accuracy as optimization metric.
The online benchmarking is, however, inherently imbalanced because we have longer non-stimulus than stimulus phases (see Fig.~\ref{fig:online_results}).
Hence, we report precision and recall.
Reliance on a legacy dataset required separate models for temperature and ozone classification (negative result for simultaneous classification not shown).
\begin{table}[t]
    \centering
    \caption{Classification results of best found models on the quasi-online simulation.}
    \label{tab:classification_results}
    \begin{tabular}{@{}p{1.2cm}p{1.2cm}p{1.5cm}p{1.2cm}p{1.2cm}@{}}
    \toprule
        Stimulus & Preprocessing & Precision & Recall & Accuracy \\\midrule
         \multirow{3}{0.75cm}{Temperature} & AMM & \textbf{0.341} & 0.501 & 0.786 \\
         & SMM & 0.324 & \textbf{0.659} & 0.750 \\
         & RDS & 0.267 & 0.234 & \textbf{0.795} \\
     \bottomrule
    \end{tabular}
    \vspace{-15pt}
\end{table}

\subsection{Deep learning model selection}
\label{sec:model_selection}
\noindent We performed automatic hyperparameter optimization using Optuna, 50 trials each. 
The best heat stimulus models have a validation accuracy of 85.8\% (size: 69,264~Byte) for AMM, 84.6\% (size: 63,952~Byte) for SMM, and 84.6\% (size: 64,592~Byte) for RDS. 
The best ozone stimulus models have a validation accuracy of 83.2\% (size: 78,864~Byte) for AMM, 72.7\% (size: 82,064~Byte) for SMM, and 87.7\% (size: 60,752~Byte) for RDS (details on hyperparameters:~\cite{Heck2025}). 

Using the best models based on our training data, we conducted quasi-online simulations for temperature by replaying pre-recorded, unseen data under conditions that emulate an actual online experiment. 
Results are reported in Table~\ref{tab:classification_results}. 
The results aggregate the two measurement channels at each plant by generating individual predictions for each channel and taking the minimum of the two. 
The sensor considers a stimulus to be detected if the minimal prediction exceeds~0.5.
In simulation, aggregating predictions via the minimum yielded better results than using the maximum or mean (78.6\% vs. 45.9\% or 68.2\% accuracy in case of AMM). 
Simulation performance slightly declined, as expected due to the modified task and use of unseen data. 
For our real online experiments, we selected the preprocessing method AMM, as it achieved the highest precision, while maintaining accuracy comparable to RDS.
Due to data scarcity, we simulated temperature experiments and used AMM for the real online ozone experiments.

\subsection{Online classification}
\label{sec:online_classification}
\noindent The online classification accuracies using AMM and minimum aggregation are 78.1\% for the heat and 46.7\% for the ozone stimulus.
We display six example stimuli applications (4~heat; 2~ozone) in Fig.~\ref{fig:online_results}.
Heat detection matches simulation accuracy, while ozone detection underperforms due to overdetecting, likely caused by limited training data with altered electrode placement (see~\cite{Aust2025}).
Although the plant's EDP reacts within seconds, the bio-hybrid sensor's effective response time depends on how much of the 10-min input window the model (predicting every 10~s) requires for correct detection.

Even though we selected the stimuli thresholds to maximize the likelihood of detectability (literature suggests that lower stimuli amplitudes can be detected, e.g., temperature change of 2.3~$^{\circ}$C~\cite{Buss2023}, and ozone concentrations of 50~ppb~\cite{Dolfi2015}), the precision and recall in the online experiment varied notably across plants and days. 
For temperature detection our model showed precisions between 0.197 and 0.681 and recall between 0.199 and 0.781.
For ozone detection, precision ranged from 0.194 to 0.323 and recall from 0.522 to 0.981.
Our temperature model generally shows good sensitivity, but depending on the plant and time, it often produces false positives. 
The ozone model detects most ozone applications, indicating high sensitivity (recall), but this comes at the cost of an increased false positive rate, resulting in lower precision. 
Despite lower precision and recall compared to off-the-shelf sensing devices, our approach shows that it is possible to train and deploy a deep neural network on our low-cost \textit{PhytoNode} platform for autonomous online detection of two environmental stimuli using the bio-hybrid sensor.
The comparison between temperature and ozone results suggests that ozone performance can be improved with additional training data. 
This is an important step towards building a bio-hybrid sensor device capable of serving as an early warning system for air pollution or enable precision agriculture by using the crop itself as sensor.

\section{Conclusions and Future Work}

\noindent This work integrates embedded DL into our \textit{PhytoNode} to build a fully autonomous sensing platform of online, on-device phytosensing which directly indicates increased temperature or ozone levels in in-lab settings. 
Results show that heat was detected accurately, while ozone level detection was comparatively lower due to a smaller dataset used for hyperparameter optimization and training, as well as by experiment setup changes during data gathering. 
We advance the paradigm of phytosensing by demonstrating a generalizable data-driven approach with potential to adapt to a wide range of environmental factors, enabling applications across various domains. 
Future work will address single-model multi-stimulus classification, and generalizability to other species by broader data collection.

\section*{Acknowledgment}

\noindent This work is supported by EU H2020 FET project WatchPlant, grant agreement No.~101017899.


\bibliographystyle{IEEEtran}
\bibliography{IEEEabrv,main}

\end{document}